\documentclass[prc,aps,twocolumn,showpacs]{revtex4}
\usepackage{graphicx}
\begin{document}
\newcommand{\etal}{{\it et al.} }

\title{ Comment on The Evidence for a Pentaquark and Kinematic Reflections}

\author{K.~Hicks$^1$, V.~Burkert$^2$, A.~E.~Kudryavtsev$^{3,4}$, 
I.~I.~Strakovsky$^4$, S.~Stepanyan$^2$
\vspace*{0.1in}
}

\affiliation{
$^1$Department of Physics and Astronomy, Ohio University,
    Athens, OH 45701, USA\\
$^2$Thomas Jefferson National Accelerator Facility,
    12000 Jefferson Ave., Newport News, VA 23606, USA\\
$^3$Institute of Theoretical and Experimental Physics,
    25 Bolshaya~Cheremushkinskaya Street, Moscow 117259, 
    Russia \\
$^4$Center for Nuclear Studies, Department of Physics,
    The George Washington University, Washington, D.C.
    20052, USA \\
}

\maketitle

In a recent article, Dzierba \etal \cite{dzierba} discuss the 
possibility that kinematic reflections, coupled with 
statistical fluctuations, could mimic the evidence in some 
photoproduction experiments for the $\Theta^+$ resonance.  
This is, at first glance, a reasonable criticism where the 
experimental data can be confronted quantitatively 
with calculation.  However, at a deeper level of examination, there 
are considerable model assumptions that Ref.~\cite{dzierba} has
made which warrant closer scrutiny.  We shall show that it is 
doubtful that kinematic reflections of the type given in 
Ref.~\cite{dzierba} can quantitatively account for the pentaquark 
peak seen in photoproduction experiments such as 
Stepanyan \etal \cite{stepanyan} or Nakano \etal \cite{nakano}.

The mechanism suggested in Ref.~\cite{dzierba} is photoproduction 
of neutral $f_2^0$(1270) and $a_2^0$(1320) tensor mesons  
(with subsequent decay into $K^+K^-$) 
for the reaction $\gamma d \to K^+ K^- p n$,  to produce a broad 
enhancement in the mass spectrum of the $nK^+$ system. 
Although their calculation cannot reproduce the narrow peak shown in 
Ref.~\cite{stepanyan}, the statistics are low and fluctuations of 
this broad enhancement might result in a false narrow structure.  
There is no estimate in Ref.~\cite{dzierba} 
for the probability of such a fluctuation, assuming their model 
is correct.

The reaction model in Ref.~\cite{dzierba} uses 
$t$-channel Regge exchange amplitudes which has been studied
for charged $a_2^+$ photoproduction (see Ref. 10 of their 
paper).  They use the same model for neutral $a_2^0$ and 
$f_2^0$ photoproduction, and compare their calculation to 
the data of Ref.~\cite{stepanyan}, as described above. 
The exchange particle in this model is the pion (and its 
higher-mass partners on the Regge trajectory line). 
However, the neutral $\pi^0$ itself cannot participate in this 
trajectory, since this vertex is forbidden by C-parity 
(the $\pi^0$, $f_2^0$ and $a_2^0$ all have positive C-parity, 
whereas the photon has negative C-parity).  The fact that the 
lowest mass exchange particle cannot participate drastically 
alters the Regge amplitudes.  For example, in charged $a_2^+$ 
production, the pion pole is known to dominate the Regge amplitudes 
at photon energies of a few GeV.  When the pion amplitude is 
dropped, the calculated cross section for $a_2^0$ and $f_2^0$ 
production is lower by over an order of magnitude \cite{titov,lee}.

It is interesting to note that Ref.~\cite{dzierba} does not 
take the resonance parameters from charged $a_2^+$ photoproduction. 
In fact, they fit the parameters in their model to the $nK^+$ 
spectrum of Ref.~\cite{stepanyan}. Can we trust these 
exchange resonance parameters? 
One serious concern is that Ref.~\cite{dzierba} did not take 
into account the detector acceptance.  The parameters extracted 
from fitting the raw spectrum of Ref.~\cite{stepanyan} without 
acceptance corrections are unlikely to represent the underlying 
physics correctly.
In a predictive calculation, these parameters would be fit 
to an independent data set for photoproduction of $f_2^0$ and $a_2^0$ 
production, as extracted from the $K^+K^-$ decay spectrum, 
and then applied to the $nK^+$ spectrum of Ref.~\cite{stepanyan}.  

The most curious aspect of the calculation in Ref.~\cite{dzierba} is the 
value of the cross section.  As shown in Fig. 4 of Ref.~\cite{dzierba}, 
the peak in the calculation near 1.25 GeV is due to the $f_2$(1270) 
and the $a_2$(1320).  The broad background is presumably due to 
nonresonant $P$-wave production.  Again, the parameters have 
been fit to the $nK^+$ spectrum, and not the $K^+K^-$ spectrum. 
This is odd, considering that the $a_2$ and $f_2$ decay into 
$K^+K^-$ whereas the $nK^+$ spectrum depends on many factors 
(including the Fermi momentum).  Clearly, if the 
authors of Ref.~\cite{dzierba} had fit the $K^+K^-$ spectrum instead, 
the contribution of $a_2$ and $f_2$ would be much reduced.  

Furthermore, the reaction mechanism used in Ref.~\cite{dzierba} assumes that 
some helicity states of the tensor mesons are preferentially excited.  
In essence, their model assumes that the tensor mesons are polarized, 
even though the initial photon beam (and the target) are unpolarized.  
The population of particular $m$-substates of the tensor mesons results 
in an angular distribution which, when integrated over all space, 
gives a minimum which is reflected in the $(nK^+)$ mass spectrum.  
However, it has not been shown by any data to our knowledge 
that the $a_2$ and $f_2$ tensor mesons are produced in a 
polarized state in the helicity frame, as Ref.~\cite{dzierba} assumes 
when using a nucleon-flip pion exchange reaction mechanism.  
Without this untested model assumption, the broad peaks in 
the $nK^+$ mass spectrum  of Ref.~\cite{dzierba} go away.  

We wish to state that, {\it apriori}, we do not reject the 
suggestion that kinematic reflections exist, nor do we dismiss 
the possibility that the peak seen in Ref.~\cite{stepanyan} 
might be due to kinematic reflections.  The point is that 
the $a_2^0$ and $f_2^0$ are not likely candidates for this.
In any case, the $K^+K^-$ invariant mass spectrum is a much 
better starting point to determine these contributions, rather than 
fitting the $nK^+$ spectrum as was done in Ref.~\cite{dzierba}.
Finally, reaction mechanisms that depend on preferential production 
of polarized $m$-substates from unpolarized beam and target 
should be confronted with data.

To conclude, the calculated mass spectrum of Ref.~\cite{dzierba} 
is not likely to survive a careful quantitative comparison with the 
data of Ref.~\cite{stepanyan}.  At the very least, one can say that 
there is considerable model-dependence in Ref.~\cite{dzierba}.  
We believe it is fair to question the authors of Ref.~\cite{dzierba} 
for the validity of their conclusions, especially considering 
that the dominant pion pole cannot contribute 
to their Regge amplitudes.

We acknowledge useful comments from Ya. I. Azimov regarding 
C-parity conservation at the photon vertex.


\end{document}